\documentclass[a4paper]{jpconf}
\usepackage{graphicx}

\begin{document}

\title{Test of beta and antineutrino spectra symmetry in beta-decay}

\author{S. V. Silaeva and V. V. Sinev}

\address{Institute for Nuclear Research of Russian Academy of Sciences, Moscow, Russia}

\ead{vsinev@inr.ru}

\begin{abstract}
The mechanism of beta decay in nature is not understood yet. Mirrored energy spectra of electron and antineutrino can clarify the situation. A special experiment is needed to measure antineutrino spectrum from known beta-decaying isotope to compare it with the beta one. One of ongoing experiments with large volume detector can be chosen to make the experiment. Another possibility is to make a special experiment close to a powerful source of mixture of known beta-decaying isotopes. If sufficient differences in shape will be observed the method of antineutrino spectrum calculation should be revised.
\end{abstract}

\section{Introduction}

In beta decay one nucleus is transformed into another one with the emission of electron (positron) and antineutrino (neutrino). Also a nucleus can capture one of own atomic electrons from electron shell and emit monoenergetic neutrino.
\begin{equation}
{\beta}^{-}: \quad (A, Z) \rightarrow (A, Z+1) + e^{-} + \bar{\nu_e}, \quad Q_{\beta^-} = M_p - M_d
\end{equation}
\begin{equation}
{\beta}^{+}: \quad (A, Z) \rightarrow (A, Z-1) + e^{+} + \nu_{e}, \quad Q_{\beta^+} = M_p - M_d - 2m_e
\end{equation}
\begin{equation}
\textrm{Electron capture (EC)}: \quad (A, Z) + e^{-}  \rightarrow (A, Z-1) + \nu_{e}, \quad E_{\nu} = Q_{\beta},
\end{equation}
where $M_p$ and $M_d$ masses of parent and daughter nuclei.

Where inside the nucleus electron and antineutrino are born is unknown. E. Fermi created theory of beta-decay \cite{fermi} and postulated that electron and antineutrino are born together exactly in the moment of beta-decay. 
Electron and antineutrino energy spectra are mirrored to each other relatively to the point in the middle of the spectrum ($T_{0}/2$, $T_{0} \approx Q_{\beta}$). 

We can regard beta decay from another point of view. Electron is a particle that massive and charged in contrast with antineutrino that is neutral and uncharged. It is understood why electron spectrum while transforming by nucleus electric field becomes softer $-$ it passes through Coulomb field of daughter nucleus and interacts with it. But antineutrino should escape from a nucleus without any visible interaction due to the absence of charge and magnetic moment.

Antineutrino spectrum from individual isotope was never measured up to now. Only the spectrum containing thousands of individual spectra was measured in experiments with reactor antineutrinos (see for example last high statistical experiments \cite{daya}$-$\cite{double}). But using this spectrum is impossible to check if antineutrino spectrum is exactly mirrored relatively to the electron one.

The test on the mirror symmetry of antineutrino spectrum and electron one can be done by measuring antineutrino spectrum directly using any known beta beta-decaying isotope and comparing it with the beta one. Modern detector techniques allows to measure antineutrino spectrum from some individual isotope. The progress in antineutrino detection in experiments with reactor antineutrino spectrum makes it possible to measure individual isotopes antineutrinos. The problem is only to get a powerful antineutrino source of pure beta-decaying isotope.

We propose to make an experiment on measuring antineutrino spectrum using high activity radioactive $^{90}$Sr$-^{90}$Y source (1$-$2 MCi) that can be placed in the vicinity of existing now detectors (on the top) of ongoing large detector experiments (KamLAND, SNO+). The another possibility to measure known antineutrino spectrum is to install relatively small detector close to spent fuel storage of some Nuclear Power Plant (NPP).

\section{Beta and antineutrino spectra shapes}

Beta spectrum shape $P_{e}(E_{\bar{\nu}},E_{0},Z)$ can be written as
\begin{equation}
P_{e}(E_{e},E_{0},Z) = K\cdot p_{e}E_{e}\cdot (E_{0}-E_{e})^2\cdot F(Z,E_{e})\cdot C(Z,E_{e})\cdot (1+\delta(Z,A,E_{e})),
\end{equation}
where $K -$ is normalization factor, $p_{e}$ and $E_{e} -$ are momentum and energy of electron, $F(Z,E_{e}) -$ Fermi function accounting Coulomb field of the daughter nucleus, $C(Z,E_{e}) -$ the factor accounting momentum dependence of nucleus matrix element and $\delta(Z,A,E_{e}) -$ is correction factor to the spectrum shape.

The antineutrino spectrum shape $P_{\bar{\nu}}(E_{\bar{\nu}},E_{0},Z)$ can be expressed the same way as beta spectrum shape by changing $E_e$ by $E_0 - E_{\bar{\nu}}$. 

It is accounted that electron and antineutrino are born in the same moment in the nucleus and therefore all corrections should be applied to the electron and the same way (mirrored) to the antineutrino. But why? In reality we do not know how electron and antineutrino appear inside a nucleus, what is just understood that they appear simultaneously somewhere inside a nucleus and leave it in opposite directions. Electron moves in the direction opposite to nucleus spin \cite{wu}. Electron and antineutrino share energy $Q_{\beta}$ of parent nucleus relatively to the daughter one. The energy $Q_{\beta}$ can be calculated as a difference between atomic masses of parent and daughter atoms what is equivalent to difference of mass excesses. Mass excess is difference between nucleus mass and sum of nucleons masses.
\begin{equation}
Q_{\beta} = \Delta M_{parent} - \Delta M_{daughter},
\end{equation}
where $\Delta M_{parent}$ and $\Delta M_{daughter} -$ are mass excesses for parent and daughter nuclei.
$Q_{\beta}$ determines maximal electron (antineutrino) kinetic energy $T_0$. For example, for $^{90}$Y we have $\Delta M_{^{90}{\rm Y}}$ = -86494.1 keV and for the daughter nucleus $^{90}$Zr $\Delta M_{^{90}{\rm Zr}}$ = -88772.5 keV. The difference is 2278.4 keV what is exactly maximal energy of $^{90}$Y electron spectrum.

\begin{figure}
\begin{center}
\includegraphics[width=12cm]{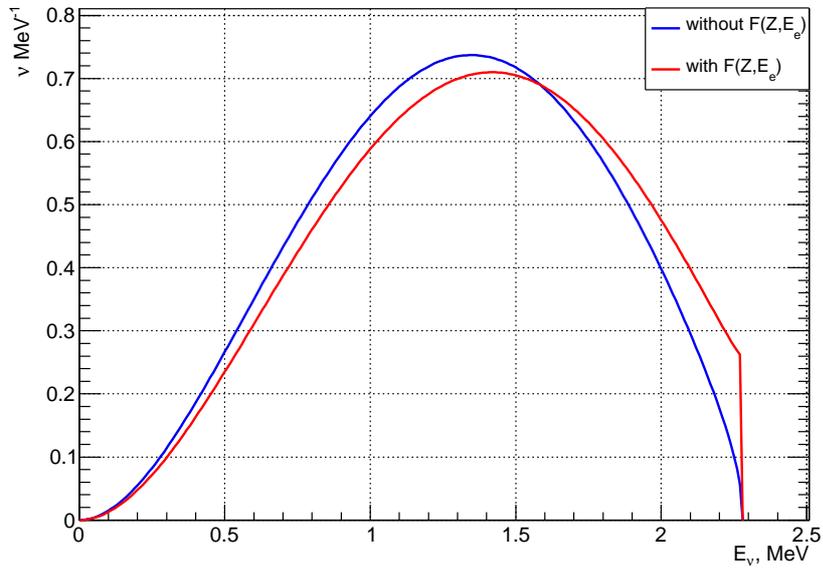}
\end{center}
\caption{\label{figone} Antineutrino spectrum from $^{90}$Y without using correction factor for electrons $F(Z,E_e)$ (blue line) and with application of $F(Z,E_e)$ (red line). }
\end{figure}

At figure \ref{figone} one can see calculated antineutrino spectrum from $^{90}$Y. There are shown two spectra: one spectrum was calculated using all corrected factors equal to unit and the other one used Fermi function $F(Z,E_e)$ that was applied to calculated spectrum. Other factors were not taken into account. We note that when using (4) with $F(Z,E_e)$ the spectrum has sharp edge at $T_0$. In case of $F(Z,E_e)$ = 1 spectrum goes to "0" smoothly.

\section{$^{90}$Sr$-^{90}$Y source for large detectors in ongoing experiments}

To measure individual isotope antineutrino spectrum one needs a powerful enough source. The using of powerful ($\sim$ 10 MCi) $^{90}$Sr$-^{90}$Y source for testing of Standard Model and looking for sterile neutrinos was proposed in \cite{sinev}. The sources in 200$-$300 kCi activity are using as electric power suppliers for some needs in hardly reached places.

\begin{figure}
\begin{center}
\includegraphics[width=12cm]{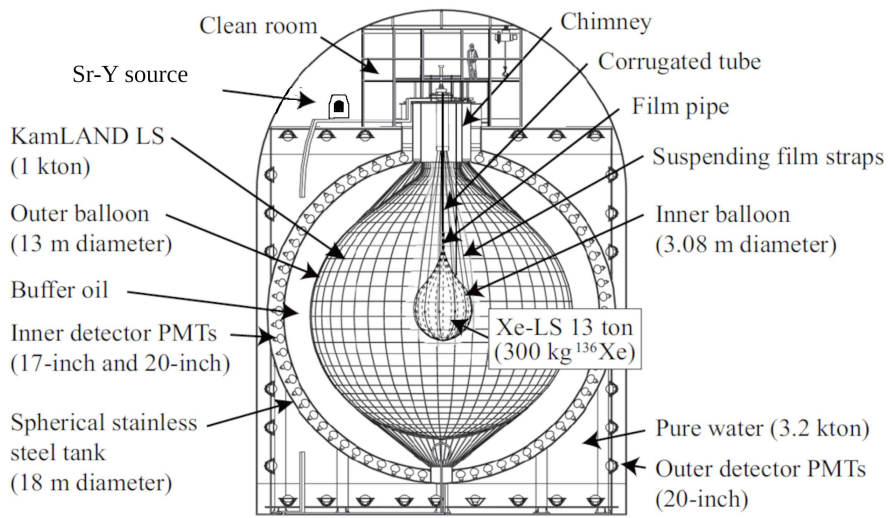}
\end{center}
\caption{\label{figtwo} Possible position of $^{90}$Sr$-^{90}$Y source on the top of KamLAND detector.}
\end{figure}

\begin{figure}
\begin{center}
\includegraphics[width=12cm]{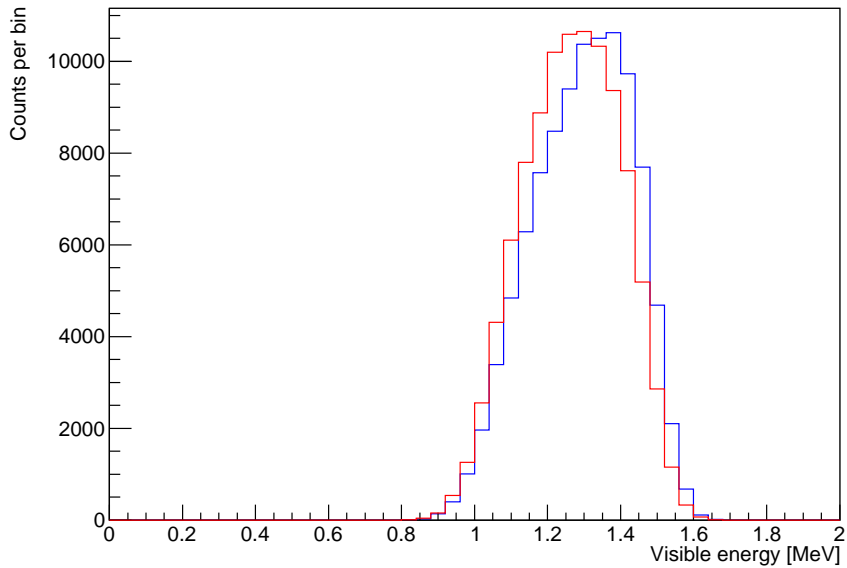}
\end{center}
\caption{\label{figthre} $^{90}$Y antineutrino spectrum that can be observed in a detector with the use of $^{90}$Sr$-^{90}$Y source 1 MCi by KamLAND during 3 years of data taking. Blue line $-$ spectrum with application of Fermi function $F(A,E_e)$, red line $-$ the spectrum without Fermi function. Energy resolution 200 ph.e. per MeV.} 
\end{figure}

An array of sources in 200-300 kCi each (1$-$2 MCi in total) can be placed on the top of detectors of ongoing experiments (KamLAND, SNO+). Scheme of the experiment is shown on figure \ref{figtwo}. For the example we choose KamLAND detector \cite{kaml}. We have calculated expected effect from 1 MCi $^{90}$Sr$-^{90}$Y source at a distance 13 m from the KamLAND detector center. At figure \ref{figthre} the spectrum from $^{90}$Sr$-^{90}$Y source, that can be observed in a detector is presented. Right side of the spectrum should be pure energy resolution of the detector in case of sharp spectrum edge. If it smoothly goes to zero the shape will be different what is seen at the figure. At the figure the difference $\sim$100$-$150 keV in maximum positions can be seen as well as difference in right slope.

Statistics that can be achieved in one year with 1 MCi source in KamLAND detector is $\sim$26700 events/year in case of sharp spectrum edge and $\sim$18700 events/year in case of smooth spectrum. If to measure 4$-$5 years the statistics can exceed 10$^5$ events.

The same antineutrino source can be applied also at SNO+ detector \cite{snop}. But the distance to the center is appeared slightly larger ($\sim$14 m) than that for the KamLAND detector. Correspondingly the events number becomes less, $\sim$23000 events/year for the sharp spectrum edge.

The application of the $^{90}$Sr$-^{90}$Y source at large scintillation detectors is also useful for other goals, not only for testing the shape of antineutrino spectrum. In \cite{sinev} was shown how this source can be used for Standard Model testing and measuring of sin$^{2}\theta_{W}$ with high accuracy. If to measure in one experiment recoil (with the same antineutrino source) recoil electrons spectrum from ($\bar{\nu_e},e^{-}$)-reaction and cross section of inverse beta decay reaction (IBD), the recoil electrons spectrum normalized on IBD cross section will not depend on the source parameters but only on sin$^2\theta_{W}$.
\begin{equation}
\frac{^{W}S(T)}{N_{\nu p}} = \frac{g^{2}_{F}}{G^{2}_{V} + G^{2}_A} \cdot F(T,\textrm{sin}^2 \theta_{W})\cdot \frac{n_e}{n_p},
\end{equation}
where $^{W}S(T) -$ recoil electrons spectrum, $N_{\nu p} -$ IBD events number, $T -$ recoil electrons kinetic energy, $F(T,\textrm{sin}^2 \theta_{W}) -$ normalized on IBD events recoil electron spectrum and $n_e$ and $n_p -$ electron and proton densities in the scintillator. 

With statistics more than 10$^5$ events in recoil electrons spectrum sin$^2 \theta_{W}$ accuracy could be less than 1\%. It is important to stress that the Weinberg angle value can be measured at very low energy in compare with accelerator experiment. If KamLAND or SNO+ detectors can see recoil electrons from antineutrino-electron scattering the test of Standard Model can be done.

\section{Small detector in vicinity of spent fuel storage}

The test of antineutrino energy spectrum shape can be also made by using a small detector ($\sim$1 $-$ 3 m$^3$) placed in some room under the reactor spent fuel storage (water pool) at a distance less than 20 m from the center of the storage. If there is no room in the same building, the special tunnel can be done that goes under the storage building. 

At NPP the spent fuel storage is placed in neighboring to the reactor building. It is water pool where spent fuel assemblies are placed after the use during 3 years in a reactor core. Normally it contains a number of assemblies accumulated in 5 years, sometimes more. Fuel amount in spent fuel storage corresponds to 1.7 reactor cores as minimum. At Neutrino-2020 Double Chozz collaboration presented the poster \cite{cecile} where the results of measuring residual neutrinos emission from two cores were shown. The registered by near detector $\sim$80 events during $\sim$20 days when both reactor cores were off (one can summarize four points at the plot for near detector). It is $\sim$4 events per day for a distance about 400 m to reactor core and detector volume 30 m$^3$. If we assume roughly the equal input in this counting rate from both stopping cores and spent fuel storages we get $\sim$1 event per day.

\begin{figure}
\begin{center}
\includegraphics[width=12cm]{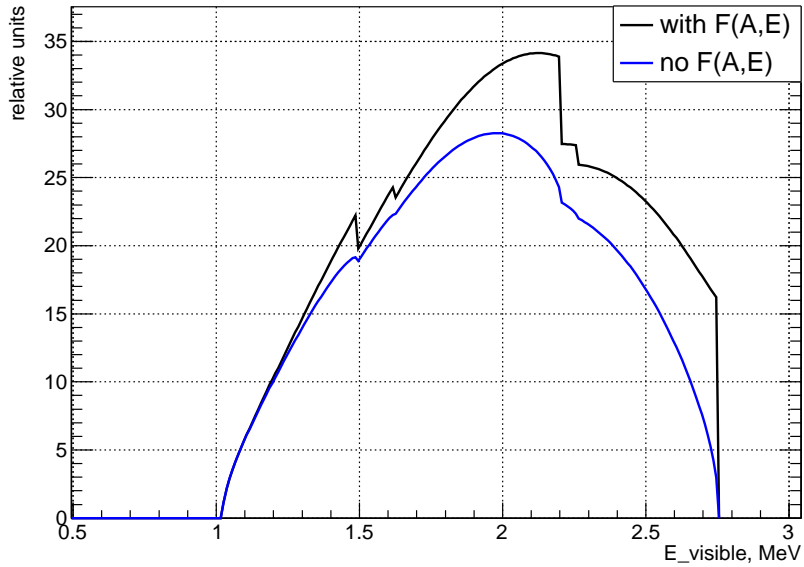}
\end{center}
\caption{\label{figfour} Spent fuel storage antineutrino spectrum as a mixture of the spectra produced by $^{90}$Y, $^{144}$Pr and $^{106}$Rh. Black line with application of $F(A,E_e)$ function to antineutrino spectrum, blue line without using $F(A,E_e)$.} 
\end{figure}

Antineutrino emission from spent fuel storage (and stopping reactor core after 3 days) is determined mainly by three nuclei: $^{90}$Y, $^{144}$Pr and $^{106}$Rh. At figure \ref{figfour} spent fuel spectrum is shown. It is calculated by summing the spectra with weights of parts of nuclei numbers for these isotopes. 

We can recalculate 1 event per day for 30 m$^3$ to a small detector 1 m$^3$ placed at 15 m from the center of spent fuel pool. Then we have $\sim$15.5 events per day or 5600 events per year. This spectrum will be seen with large background of the core ($\sim$700 events per day if a core at a distance 60 m). 

The effect is about 2\% of the background but it is concentrated in energy region below 3.5 MeV of antineutrino energy. So, it can be seen in a similar way as geoneutrino signal in KamLAND detector.
Reactor antineutrino spectrum is well known, can be easily fitted and then be subtracted from measured data.

\begin{figure}
\begin{center}
\includegraphics[width=12cm]{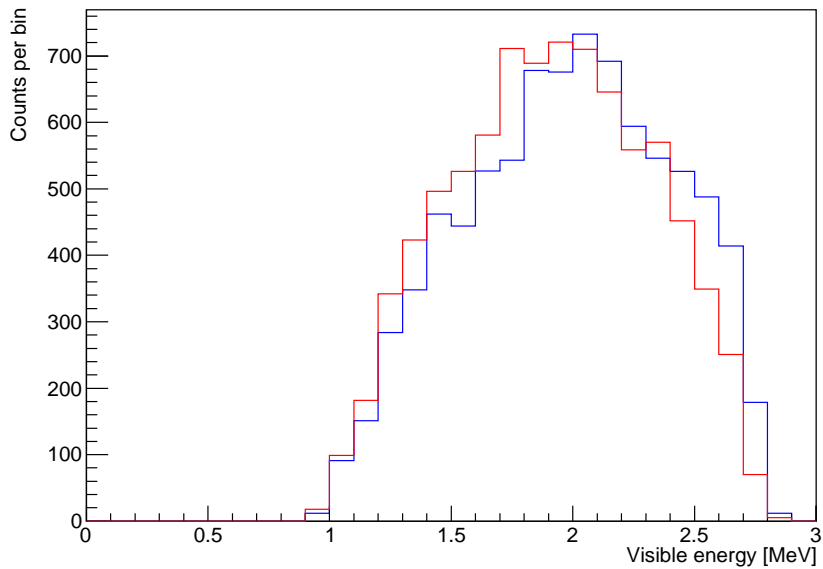}
\end{center}
\caption{\label{figfive} Spent fuel storage antineutrino spectrum simulated for the detector energy resolution (200 ph.e. per 1 MeV). Blue line with $F(A,E_e)$ function and red line without $F(A,E_e)$. Energy resolution 200 ph.e. per MeV.} 
\end{figure}

At figure \ref{figfive} we show how spent fuel spectrum looks like in case of finite energy resolution at statistics 8500 events (1.5 years of data taking). Here we also see by eye the difference in spectra in spite of energy resolution. 

\section{Conclusion}

In this work we propose to check if antineutrino spectrum is really mirror symmetric to the beta one. There may be different experiments done to check the symmetry of antineutrino and beta spectra from one isotope. 

We propose to make an experiment with artificial radioactive source $^{90}$Sr$-^{90}$Y. Using this source of 1$-$2 MCi activity placed on the top of ongoing experiments (KamLAND, SNO+) we can expect in several years to reach appropriate statistics to make a conclusion on antineutrino spectrum shape. The experiment with $^{90}$Sr$-^{90}$Y source is convenient also for testing Standard Model. Weinberg angle can be measured with high accuracy as shown in \cite{sinev}.

Test of antineutrino energy spectrum shape can be also done with a small detector at NPP if to place a detector under spent fuel pool building. The background of antineutrino from reactor core can be subtracted. In several years of measuring necessary statistics can be accumulated.

\section*{Acknowledgements}

We would like to thank L. B. Bezrukov for fruitful discussions and valuable remarks.

\end{document}